# Maximizing the Power of Principal Components Analysis of Correlated Phenotypes in Genome-wide Association Studies.

Hugues Aschard[1], Bjarni J. Vilhjálmsson[1], Nicolas Greliche[2], Pierre-Emmanuel Morange[3], David-Alexandre Trégouët[2], Peter Kraft[1]

[1] Harvard School of Public Health, department of Epidemiology, Boston, USA
[2] INSERM UMR_S 937, ICAN Institute for Cardiometabolism And Nutrition, Pierre et Marie Curie University, Paris 6, France.
[3] INSER UMR_S 1062, Aix-Marseille Université, Marseille F-13385, France.

## Abstract

Principal Component analysis (PCA) is a useful statistical technique that is commonly used for multivariate analysis of correlated variables. It is usually applied as a dimension reduction method: the top principal components (PCs) explaining most of total variance are tested for association with a predictor of interest, and the remaining PCs are ignored. This strategy has been widely applied in genetic epidemiology, however some of its aspects are not well appreciated in the context of single nucleotide polymorphisms (SNPs) association testing. In this study, we review the theoretical basis of PCA and its behavior when testing for association between a SNP and two correlated traits under various scenarios. We then evaluate with simulations the power of several different PCA-based strategies when analyzing up to 100 correlated traits. We show that contrary to widespread practice that testing the top PCs only can be dramatically underpowered since PCs explaining a low amount of the total phenotypic variance can harbor substantial genetic associations. Furthermore, we demonstrate that PC-based strategies that use all PCs have great potential to detect negatively pleiotropic genetic variants (e.g. variants with opposite effects on positively correlated traits) and genetic variants that are exclusively associated with a single trait, but only achieve a moderate gain in power to detect positive pleiotropic genetic loci. Finally, the genome-wide association study of five correlated coagulation traits in 685 subjects from the MARTHA study confirms these results. The joint analysis of the five PCs from the coagulation traits identified two new candidate SNPs, which were most strongly associated with the 5$^{th}$ PC that explained the smallest amount of phenotypic variance.

# Introduction

Common diseases such as asthma or type-2 diabetes are often studied through related phenotypes. The identification of genetic variants that influence these correlated, and potentially intermediate, phenotypes may hold the key to understanding the genetic architecture of the disease in question. While many studies analyze each of these phenotypes separately, their joint analysis is likely to have increased statistical power in many instances[1]. However, integrating association signals observed between a single SNP and multiple traits in a single comprehensive framework is not always straightforward. Simple approaches, such as Fisher's method applied to univariate analysis of each phenotype can yield inflated type I error rate when the traits are correlated. More complex methods can indeed be applied. For example, mixed models have been proposed recently for modeling the covariance structure caused by correlated phenotypes as well as population structure [2,3], but this approach quickly becomes computationally intractable as the number of phenotypes grows. Several other sophisticated approaches for jointly analyzing correlated traits exist, but their respective advantages and limitations are generally not well understood. Many of them are related to canonical correlation analysis (CCA) [4,5], and consist in identifying the linear combination of a set of variables that is the most highly correlated with any linear combination of a second set of variables (equivalent to a one-way MANOVA when analyzing a single SNP). A slightly different strategy, also related to CCA, consists in focusing first on the core structure of the data by extracting the shared and specific parts of the original phenotypes. The guiding principle here is to build new variates using spectral decomposition of either the covariance or correlation of the original phenotypes, as done in principal component analysis (PCA), so that these new variates, or components, are uncorrelated from each other and explain successively maximum variance. There are two major advantages in applying such a technique. First, because all components are independents, their multivariate analysis is straightforward. Second, since most of the phenotypic variance is usually captured by the first few components, this approach can be used as a variable reduction method.

In this study we focus on PCA-based strategies for multivariate analysis of multiple traits, a common application in genetic epidemiology that has been proposed for linkage analysis (e.g. [6]) and genome-wide association studies (GWAS)[7-13]. For the purpose of GWAS, and following the principle of dimension reduction, one typically tests for an association between each SNP and the first few principal components (PCs) that explain most of the total phenotypic variance. For example, in a recent GWAS of facial morphology, researchers used the first 11 PCs from the facial shape phenotypes, each explaining at least 1% of the total phenotypic variance [7]. In another recent GWAS of metabolic syndrome traits, PCA was applied independently on 6 domains, including a total of 19 phenotypes. In this study the few first principal components were extracted for each domain so that they explained at least 55% of the total phenotypic variance [8]. Downstream from selecting the top PCs, various different analysis strategies have been applied in the literature. In many studies these top PCs were analyzed individually (univariate analysis) and their loadings (i.e. the weight of each raw phenotypes, e.g. [9-12]) were reported. Other studies either conducted standard multivariate analysis of these components, e.g. [8]) or built more complex combined tests that accounted for other sources of information [13].

Overall, previous work has clearly demonstrated the utility of PCA-based GWAS strategies that focus on the top principal components. However questions remain about which PCA strategy is the most powerful for detecting associations, how to interpret signal observed on PCs, and under what conditions are PCA-based multivariate analysis expected to outperform a simple univariate analysis of the raw phenotypes. The aim of this study is to address these questions both analytically and using simulations. Unlike previous studies that assume and model complex associations between the SNP and the traits studied [14] or discuss causal inference [1], this study explores the use of PCA for performing GWAS in multiple correlated phenotypes simultaneously without making any strong assumption about the underlying model. First, we review some of the theoretical basis for PCA and describe the expected behaviors of univariate analysis of each PC when analyzing two correlated phenotypes. We then explore the power of various PCA-based strategies when analyzing a larger number of simulated correlated phenotypes. We show, contrary to the current prevalent belief, and as previously discussed for the analysis of non-genetic data (e.g. [15]), that principal components explaining a small amount of total phenotypic variance can be as important as those explaining large amount of variance. In many scenarios these components capture a large proportion of the genetic effect. Discarding these lower-variance PCs can, therefore, severely decrease power to detect genetic variants associated with one or more of the traits. Finally, we confirmed the usefulness of these results by analyzing 5 coagulation related quantitative traits in 685 subjects from the MARTHA study [16,17].

**Material and Methods**

*Analysis of two phenotypes*

For illustration, consider a hypothetical model where two positively correlated and normally distributed phenotypes, $Y_1$ and $Y_2$ with mean 0 and variance 1 both depend on an unknown variable $U$ and a scaled genotype $G$, that are also normally distributed with mean 0 and variance 1. Let $c$ denote the correlation between $Y_1$ and $Y_2$ due to $U$ and $v_1$ and $v_2$ the proportion of variance of $Y_1$ and $Y_2$ explained by $G$, respectively. We assume the effects of $U$ and $G$ on $Y_1$ and $Y_2$ are positive and, $(c+v_{max})$, where $v_{max}$ is the maximum of $v_1$ and $v_2$, can vary in [0, 1], so that the two traits can be expressed as:

$$\mathbf{y_1} = \sqrt{c} * \mathbf{u} + \sqrt{v_1} * \mathbf{g} + \sqrt{(1-c-v_1)} * \mathbf{\varepsilon_1} \quad (A)$$
$$\mathbf{y_2} = \sqrt{c} * \mathbf{u} + \sqrt{v_2} * \mathbf{g} + \sqrt{(1-c-v_2)} * \mathbf{\varepsilon_2} \quad (B),$$

where $\varepsilon_1$ and $\varepsilon_2$ denote independent random noises that are normally distributed with mean 0 and variance 1. The principal components of these two traits are $\mathbf{pc_1} = \frac{1}{\sqrt{2}} * (\mathbf{y_1} + \mathbf{y_2})$ and $\mathbf{pc_2} = \frac{1}{\sqrt{2}} * (\mathbf{y_1} - \mathbf{y_2})$, which can be re-written as:

$$\mathbf{pc_1} = \frac{1}{\sqrt{2}} * \left( (2*\sqrt{c}) * \mathbf{u} + (\sqrt{v_1} + \sqrt{v_2}) * \mathbf{g} + \sqrt{(1-c-v_1)} * \mathbf{\varepsilon_1} + \sqrt{(1-c-v_2)} * \mathbf{\varepsilon_2} \right) \quad (C)$$
$$\mathbf{pc_2} = \frac{1}{\sqrt{2}} * \left( (\sqrt{v_1} - \sqrt{v_2}) * \mathbf{g} + \sqrt{(1-c-v_1)} * \mathbf{\varepsilon_1} - \sqrt{(1-c-v_2)} * \mathbf{\varepsilon_2} \right), \quad (D)$$

The total phenotypic variance explained by **pc₁** and **pc₂**, and noted respectively by $s_1$ and $s_2$, can be defined as (see **text S1**):

$$s_1 = \frac{1 + c + \sqrt{v_1 v_2}}{2} \quad (E)$$

$$s_2 = \frac{1 - c - \sqrt{v_1 v_2}}{2} \quad (F)$$

From this it follows that $v_{pc1}$ and $v_{pc2}$, the proportion of variance of **pc1** and **pc2** explained by G can be respectively expressed as (see **text S1**):

$$v_{pc1} = \frac{v_1 + v_2 + 2 \times \sqrt{v_1 v_2}}{2 * (c + 1 + \sqrt{v_1 v_2})} \quad (G)$$

$$v_{pc2} = \frac{v_1 + v_2 - 2 \times \sqrt{v_1 v_2}}{2(1 - c - \sqrt{v_1 v_2})} \quad (H)$$

The power of the association tests between $G$ and, $Y_1$, $Y_2$, $PC_1$ and $PC_2$ can then be compared for different sample size and genetic effect. If $N$ is the sample size, the Wald test for association, equal to $N \times \hat{v}$, where $\hat{v}$ is the estimated proportion of variance explained by G, follows a non-central chi-square with one degree of freedom (df) and non-central parameter equal to $\delta = N \times v$, so the power equal [18]:

$$\text{Power} = 1 - F(\chi^2_{1,1-\alpha,0} | 1, \delta) \quad (I)$$

where $F(\chi^2|d,\delta)$ is the cumulative probability function of the non-central chi-square distribution with $d$ degrees of freedom and non-centrality parameter $\delta$; $\chi^2_{d,p,\delta}$ is the inverse of $F$, i.e. the quantiles of the non-central chi-square distribution, and $\alpha$ is the type I error rate. Since the two PCs are independents, one can also define a joint test of the PCs by summing the 1df non-central chi-square from each PC to form a 2df non-central chi-square. The power for such a test is equal to:

$$\text{Power} = 1 - F(\chi^2_{2,1-\alpha,0} | 2, \delta) \quad (J)$$

For simplicity, we derived the proportion of variance explained and the power to detect the effect of a genetic variant under the assumption that the effect of the unknown source of correlation, *U*, and the genetic effects on the traits were all positive (and so in the same direction). All other patterns where the genetic effects are consistent with the correlation (e.g. negative correlation and genetic effect in opposite direction on the two traits) produce the exact same results. The extension to the situation where the genetic effect is opposite to the correlation is straightforward.

*Analysis of five correlated phenotypes*

We simulated four series of 10,000 dataset of 5,000 subjects, each series corresponding to a given model of correlation pattern. For each subject we generated a SNP *G* with minor allele frequency of 0.3 and five phenotypes $Y_1$ to $Y_5$. The phenotypes were generated as defined in equation (A), and *c* was defined so that the average correlation pattern between the *Y*s match the following correlation matrices:

$$model1 = \begin{pmatrix} 1 & 0.50 & 0.31 & 0.15 & 0.07 \\ 0.50 & 1 & 0.31 & 0.15 & 0.15 \\ 0.31 & 0.31 & 1 & 0.09 & 0.04 \\ 0.15 & 0.15 & 0.09 & 1 & 0.02 \\ 0.07 & 0.07 & 0.04 & 0.02 & 1 \end{pmatrix}, \quad model2 = \begin{pmatrix} 1 & 0.80 & 0.63 & 0.32 & 0.09 \\ 0.80 & 1 & 0.63 & 0.32 & 0.09 \\ 0.63 & 0.63 & 1 & 0.09 & 0.07 \\ 0.32 & 0.32 & 0.09 & 1 & 0.03 \\ 0.09 & 0.09 & 0.07 & 0.03 & 1 \end{pmatrix}$$

$$model3 = \begin{pmatrix} 1 & 0.30 & 0.30 & 0.30 & 0.30 \\ 0.30 & 1 & 0.30 & 0.30 & 0.30 \\ 0.30 & 0.30 & 1 & 0.30 & 0.30 \\ 0.30 & 0.30 & 0.30 & 1 & 0.30 \\ 0.30 & 0.30 & 0.30 & 0.30 & 1 \end{pmatrix}, \quad model4 = \begin{pmatrix} 1 & 0.70 & 0.70 & 0.70 & 0.70 \\ 0.70 & 1 & 0.70 & 0.70 & 0.70 \\ 0.70 & 0.70 & 1 & 0.70 & 0.70 \\ 0.70 & 0.70 & 0.70 & 1 & 0.70 \\ 0.70 & 0.70 & 0.70 & 0.70 & 1 \end{pmatrix}$$

When genetic effect on a trait was simulated, the proportion of variance in that trait explained by *G* was drawn from a uniform distribution with minimum 0.001 and maximum 0.005, independently to phenotypic correlation. When pleiotropic effects were generated, the *K*=2,..,5 phenotypes affected by the SNP were randomly chosen (with equal probability). However, we also simulated situations where the pleiotropic effect of the SNP reflects the phenotypic correlation pattern, that is, assuming that the most highly correlated traits are more likely to be associated with the same genetic variants. To do so, we assigned the probability that the $i^{th}$ phenotype was associated with *G* to be proportional to its correlation with other traits. For example in the presence of a pleiotropic effect on two phenotypes under *model 2*, the two traits $Y_1$ and $Y_2$, correlated at 0.8, had much higher chances to be selected. Note that this modification does not affect the simulations results based on *model 3* and *model 4*, since all phenotypes are equally correlated to the others.

### *Combined analysis of venous thrombosis related phenotypes*

Five quantitative intermediate phenotypes for venous thrombosis risk measured in the MARTHA study were analyzed in this report for illustrative purposes. MARTHA individuals are unrelated Europeans patients with venous thromboembolism, consecutively recruited at the Thrombophilia center of La Timone Hospital (Marseille, France) between January 1994 and October 2005. All subjects were typed for genome-wide single nucleotide polymorphisms (SNPs) using the Illumina Human610-Quad and Human660W-Quad Beadchips. Five coagulation related phenotypes were studied: three plasma levels of coagulation factors, Fibrinogen (FIB) factor VIII (FVIII), and von Willebrand factor (vWF); the activated partial thromboplastin time (aPTT) and the standardized Anticoagulant response to Agkistrodon contortrix venom (ACVn). These phenotypes have already been analyzed under a GWAS framework using either raw genotypes or imputed SNPs according to HapMap 2 database. A detailed description of the cohort and the phenotypes can be found in [16,17,19]. In the current application, we further conducted an imputation analysis based on the 1000 Genome 2012-02 release using the minimac software (release 2012-03-14). SNPs with imputation quality Rsq>0.3 and minor allele frequency (MAF) >0.01 (*N* = 8,862,493) were used in this study. Imputed SNPs were tested for association with the raw phenotypes and the derived PCs using a linear regression analysis in which allele dosage of imputed SNPs was used. For this, we employed the mach2qtl software [20]. These association analyses were conducted in a sample of 685 patients with no missing phenotype information and were adjusted for age, sex, anti-coagulant therapy, smoking and the four first principal components derived from the genome-wide genotype data to handle any undetected population stratification.

## Results

*Comparison of power for the analysis of two phenotypes*

Let us first consider the analysis of two correlated traits $\mathbf{y_1}$ and $\mathbf{y_2}$ (equation (A) and (B)) where the tested SNP only has an effect on $\mathbf{y_1}$. When the genetic effect is small, so that the contribution of the SNP in the correlation between the two traits is negligible, the two PCs (equation (C) and (D)) can be approximated by:

$$\mathbf{pc_1} \approx \frac{1}{\sqrt{2}} * \left((2*\sqrt{c})*\mathbf{u} + (\sqrt{v_1})*\mathbf{g} + \sqrt{2*(1-c)}*\mathbf{\varepsilon'_1}\right) \quad (K)$$

$$\mathbf{pc_2} \approx \frac{1}{\sqrt{2}} * \left((\sqrt{v_1})*\mathbf{g} + \sqrt{2*(1-c)}*\mathbf{\varepsilon'_2}\right), \quad (L)$$

where $\mathbf{\varepsilon'}$ are normal $\mathcal{N}(0,1)$ variables. The proportion of the total variance explained by $PC_1$ and $PC_2$ (equation (E) and (F)) simplifies to $s_1 = (1+c)/2$ and $s_2 = (1-c)/2$ respectively, so that these two proportions only depend linearly on $c$, the correlation between the two traits. Hence, $s_1$ becomes approximately equal to $s_2$ when $c$ is small, but larger than $s_2$ when $c$ increases.

From equation (K) and (L) we can see that $PC_1$, although it depends slightly on $G$, captures essentially the effect of $U$, the unmeasured variable representing the shared effect. Conversely the effect of $U$ is, by construction in this example, not captured by $PC_2$, which depends only on the effect of the SNP on $Y_1$ plus some residual noise. This noise scales with the phenotypic correlation, decreasing dramatically with increasing value of $c$. This makes $\mathbf{PC_2}$ the most relevant component to capture a genetic variant that affect only a single phenotype. However, as noted above, for moderate correlation, $G$ has a non-negligible effect on both PCs, so that combining the results from the two components might be a relevant alternative to the test of $\mathbf{PC_2}$ only. **Figure 1** presents a comparison of the power for the test of association between $G$ and respectively $PC_1$, $\mathbf{PC_2}$ and the combined PCs, against the test of association between $G$ and $Y_1$ when the SNP is associated with $Y_1$ and but not with $Y_2$. In this case the combined PCs test and the test of $PC_2$ has greater power than the test of $Y_1$ when the correlation is large, but lower power when correlation is low. This gain in power tends to be larger for the combined PCs as compared to $PC_2$ when the effect of $G$ on the trait is large, or when the sample size increases (**Figure 1A**). Conversely, the test of $PC_2$ becomes optimal when sample size or genetic effect is small, and correlation is larger than 0.5 (**Figure 1B**).

When the tested SNP affects both traits, one can similarly derive the power for each of the three previous tests against the test of $Y_1$ and $Y_2$. **Figure 2** present such results for correlation $c$ equal to 0.1, 0.5 and 0.9, while assuming $G$ explains 0.5% of the variance of $Y_1$ and between 0 to 1% of the variance of $Y_2$. Furthermore, we considered the directionality of the SNP effects for the second trait: the same direction as on the first trait (positive pleiotropy), and effects in opposite directions on the two traits (negative pleiotropy). Similarly to the previous scenario, the proportion of the total variance explained by $PC_1$ increases with $c$, but conversely $PC_1$ displays most of the genetic effect if $G$ has similar effects on both of the traits (i.e. is in same direction). This can be explained by considering $v_{pc1}$ and $v_{pc2}$, the proportion of variance explained by $G$ on the PCs (equation (G) and (H)). For example, when $v_1 = v_2 = v'$, $v_{pc2}$ is null and $v_{pc1}$ can be approximated by $v' * 2/(c + 1)$. Since $c$ is by definition smaller than 1, the test of $PC_1$ in this specific case will always outperform the test of each trait independently and the test of $PC_2$. When the genetic effects are in opposite directions (negative pleiotropy), large gains in power can be achieved by testing for $PC_2$, while testing for $PC_1$ has almost no power. This is consistent with the results of Korte et al. 2012 [2], where they reported increased power to detect trait specific and negative pleiotropic genetic effects when accounting for the correlation between the traits. In this case, the effect of $G$ on $PC_2$ can be approximated by $v' * 2/(1 - c)$, so that the effect of $G$ increases exponentially with increase in $c$. Although the expected power of the combined PCs analysis in all these situations is usually lower than the power of single PCs when assuming a specific direction of genetic effect on the traits, it offers a good compromise, allowing for reasonable power without assuming specific hypothesis about the genetic effect. A summary of how the different tests based on $PC_1$, $PC_2$, and the combined test of the two PCs behave for moderate sample size, is shown in **Table 1**.

*Comparison of power for the analysis of five phenotypes*

When more than two phenotypes are considered, deriving the power of the various analysis strategies quickly becomes too complex to be comprehensively expressed since it depends on many factors, including the correlation pattern between the phenotypes and the magnitude and direction of effect of the tested SNP on each phenotype. To explore the power of univariate and multivariate tests in these situation we simulated data for five phenotypes, $Y_1$ to $Y_5$, under various scenarios and we estimated the power of four strategies: the univariate test of the original phenotypes, the univariate test of each PC, the test of the PC displaying the largest genetic effect after correcting for multiple testing, and the combined test of all PCs. We considered four correlation models across the phenotypes analyzed: 1) a gradient of moderate to low correlations, 2) a gradient of strong to low correlations, 3) a uniform moderate correlations and 4) a uniform strong correlations (see *Methods*). For each model we studied first situations where the genetic variant is associated with a single trait (**Figure 3**), and then situations where the genetic variant has a pleiotropic effect (**Figure 4**).

Generally, no single PC was optimal to detect the associated SNP over all scenarios considered. The association pattern between the SNP and the five components varied greatly across the set of parameters we used. Sometimes the strongest association was with the PCs explaining most of the variance (e.g. when the genetic effect is associated with the 5 traits and the correlation between the trait is smaller than 0.5); sometimes the strongest association was with the PCs explaining the smallest amount of the total variance (e.g. when the SNP is associated with several, but not all traits, and there is correlation >0.5 among all traits). Hence, in most situations, the strategy of testing all PCs and picking the one with the strongest signal was more efficient than focusing on the PCs explaining most of the variance. However, because most PCs showed association signals, the combined approach was on average the most powerful except when SNP had an effect on a single trait that had correlation lower than 0.5 with the other traits analyzed, or in the presence of a fully pleiotropic effect (i.e. the SNP is associated with all 5 traits) and the correlation between traits is homogeneous and small (<0.5).

In all these simulations we considered situations where the genetic effect was in the same direction for all phenotypes and we generated pleiotropic patterns independently from the phenotypic correlation patterns. As shown in **Figure S1**, allowing for effects in opposite directions dramatically increases the detection power of all three PC-based approaches (the univariate PC analysis, the combined PCs and the "best PC" approach). When the pleiotropic effect was generated to reflect the phenotypic correlation (the most highly correlated traits where more likely to be associated to the genetic variant, *see Methods*), the association patterns (**Figure S2**) were similar to those observed without accounting for the correlation among traits (**Figure 4**).

### *When a very large number of phenotypes are analyzed*

The simulations in the previous section show that when a relatively small number of phenotypes are analyzed jointly, the genetic association signal is spread across most or all of the PCs, making the analysis of a single PC or the combined analysis of a sub-group of PCs less powerful as compared to the combined analysis of all PCs. However, when the number of phenotypes becomes very large this will not be always true, since the large increase in degree of freedom will balance any small association signal displayed on each PCs.

Consider a scenario with genome-wide genetic data and a very large number of traits, say for example, circulating levels of a hundred metabolites. Under such a scenario, when analyzing 2000 subjects and assuming that the genetic effects are of the same order of magnitude as in the previous simulations (i.e. proportion of variance explained between 0.1% and 0.5%), the power of the univariate test of the raw phenotypes at genome-wide significance level (and before any correction for the hundred 100 tests conducted) is below 1%. Depending on the correlation pattern, and the level of pleiotropy, focusing on a sub-group of PCs may increase or decrease power. This is demonstrated in **Figure 5**, where the power at $5 \times 10^{-8}$ significance level is shown for different tests when analyzing 100 phenotypes simulated using a gradient of correlation from 0 to 0.9 (extended *model 2* from the previous section, but using the more complex simulation scheme SC1 described in **supplementary text S2** and **supplementary figure S3**), and generating genetic pleiotropic effects at random, i.e. assuming genetic effects can appear on any phenotype with equal probability. We constructed two series of 100 tests by combining either the smallest *n* PCs that explain the least amount of the total phenotypic variance (i.e. the smallest eigenvalues), or the largest *n* PCs that explain the largest amount of the total phenotypic variance (i.e. the largest eigenvalues), with *n* varying from 1 to 100. When the genetic variant affected 5 traits, we observed a substantial gain in power when focusing on the 10 PCs corresponding to the 10 smallest eigenvalues as compared to the combined analysis of all PCs (power was 0.59 and 0.23 respectively). However, using the same simulation scheme, that same test (10 PCs corresponding to the 10 smallest eigenvalues) was dramatically underpowered when the SNPs was associated with 20 traits (power was 0.63 and 0.99 respectively).

While the optimal strategy will highly depend on the underlying model, we observed that a naïve approach that consists in analyzing the PCs jointly in a few sub-groups based on their eigenvalues, can significantly improve power. For example in **figure S4** we show that combining the joint signal of a subset of the PCs having the highest eigenvalues with the joint signal from the remaining PCs (having the lowest eigenvalues) can have increased power in many hypothetical scenarios with a wide range of latent variables and different patterns of genetic effects (see simulation schemes SC1, SC2 and SC3 in **text S2** and **figure S3**). Applying the Fisher's method, such a test can be written as:

$$T_K = -2 * \left[ log\left(1 - F\left(\sum_{i=1}^{K} \chi_i^2 \middle| K\right)\right) + log\left(1 - F\left(\sum_{i=K+1}^{N} \chi_i^2 \middle| N - K\right)\right)\right],$$

where $K$ is the number of top PCs included in the first group, and $N$ the total number of PCs is equal to 100. The $\chi_i^2$ is the chi-square association statistic between the SNP and the $i^{th}$ PC, and $F(\chi^2|d)$ is the cumulative probability function of the central chi-square distribution with $d$ degrees of freedom. Since all PCs are independents, $T_K$ follows a chi-square distribution with 4 degree of freedom under the null hypothesis of no association with any of the PC.

### GWAS of coagulation related phenotypes

To illustrate the importance of including principal components that explain a small proportion of total phenotypic variance in the analysis we conducted a genome-wide scan of five coagulation related phenotypes in 685 individuals from the MARTHA study, namely fibrinogen (FIB), factor VIII (FVIII), von Willebrand factor (vWF), the activated partial thromboplastin time (aPTT) and the standardized anticoagulant response to Agkistrod on contortrix venom (ACV). All these phenotypes reflect global coagulation activity and display moderate to strong correlation. The correlation matrix between these traits (**supplementary table S1**) was very similar to that simulated in *model 2*, with a gradient of absolute value correlation varying between 0.75 (FVIII and vWF) to 0.013 (FIB and aPTT). The five principal components extracted from the standardized phenotypes explained 46.22, 18.86, 18.32, 12.48 and 4.11 percent of the total phenotypic variance respectively. The individual trait loadings for each of the 5 PCs are presented in **supplementary table S2.**

The results from the univariate analysis of each original trait and the univariate analysis of each PC (adjusted for multiple hypothesis testing using Bonferroni correction) compared to the multivariate analysis of the PCs were consistent with the conclusions drawn from the simulation study. All of the five loci that were found to be genome-wide significant ($P<5\times10^{-8}$) in a single trait analysis were also significant in the combined PC analysis (**Table 2**). Conversely, focusing for example on the top two PCs explaining more than 55% of the variance as done in [8] would have identified only one of these five SNPs. The combined PCs analysis furthermore identified two new variants, one on chromosome 18q21.2 located between two genes, *DCC* and the *RPS8P3* pseudo-gene ($P=1.7\times10^{-8}$). This SNP was suggestive genome-significant based on the univariate analysis of FVIII ($P=1.7\times10^{-7}$ and beta coefficient of the coded allele (allele T against C) of 1.8) and slightly associated with vWF ($P=0.035$ and beta coefficient of 0.89). The second locus found in the combined PCs analysis was just below the genome-wide significant level ($P=5.8\times10^{-8}$). It is located on chr10p11.22 lying between the interesting genes *ITGB1* and *NRP1*. This SNP had a marginal effect on FVIII and vWF (the beta coefficients of the coded allele (allele A against G) were 0.502 and 0.297, respectively) that did not rich suggestive genome-wide significance ($p$-value of $6.6 \times 10^{-7}$ and 0.0073 respectively)

Several patterns of association observed in the simulation study were also observed in the empirical data. First, the genetic signal at the most associated loci was spread out across the PCs with association pattern changing across loci; thus, focusing only on the univariate signal from the top PCs was suboptimal. Applying the combined PCs approach on the top 2 or top 3 PCs was underpowered for the same reason (data not showed). When the SNP only affected a single trait that was moderately correlated to the others, there was no gain in using PCA-based approaches (e.g. SNPs rs6025, rs710446 and rs191945075 in **Table 2**), however when the affected trait had correlation with another trait above 0.5, the combined PCs had the highest power (e.g. SNPs rs576123, rs183013917 and rs76854392 in **Table 2**). Indeed we noted that association signals at new loci identified by the combined PCs analysis were driven by signal from the last PC, which explained 4% of the total phenotypic variance. These new signals involve non-pleiotropic effects, or at least unbalance genetic effect on FVIII and vWF, the most correlated traits. This confirms: 1) that PCA-based analysis of multiple traits can improve detection of genetic variants harboring pleiotropic effects but also those affecting a single trait; and 2) that PCs explaining a low amount of variance can be as important as those explaining large amount.

**Discussion**

Principal component analysis is a common tool that has been widely used for the combined analysis of correlated phenotypes in genetic linkage and association studies. In this study we show that PC-based analyses that focus on the few components explaining most of the phenotypic variance, as done in many studies, is generally suboptimal. By deriving the power for PC analysis in a simple case of two phenotypes and by conducting simulations for more complex situations, we show that a genetic association signal may in practice be spread across many or all of the principal components. Under many realistic scenarios important genetic signals, e.g. trait specific or negative pleiotropic genetic effects (e.g. positive correlation and opposite genetic effects), are captured by the PCs explaining the least amount of the total phenotypic variance. We demonstrated that combining the signal from multiple PCs, including those explaining a small amount of variance, is an efficient strategy for the identification of genetic variants having pleiotropic effect and those with no pleiotropic effect but associated with a trait that is highly correlated with the other traits in the study.

Since PCs are linear combinations of the raw phenotypes, testing jointly the genetic effect on all PCs is asymptotically similar to applying the Fisher's method when the traits analyzed are uncorrelated. More generally, it is closely related to standard multivariate approaches such as one-way MANOVA. The power of the joint PC test is also quite similar to that of the analytic approach implemented in the software MultiPhen, which applies ordinal logistic regression using the genetic variant as the outcome and the phenotypes as predictors [21](**Figure S5**). However using principal components may be easier to implement in a genome-wide analysis context and offer more flexibility than these integrated approaches. For example, it is straightforward to apply alternative tests for association that capture non-linear effects to each of the PCs, and then combine the results of these tests across PCs using Fisher's product test. This approach may be particularly useful in the presence of heterogeneous genetic effects, as in the case of gene-gene or gene-environment interactions with an unknown (or unmeasured) effect modifier [22]. Moreover, we noticed that Multiphen has an inflated type I error rate when analyzing many correlated phenotypes in moderate sample size (e.g. 50 phenotypes in 1000 subjects), while combining the PCs remains robust at least among scenario we considered (**Figure S6**).

Although focusing only on a few PCs is unlikely to improve statistical power when the total number of phenotypes analyzed jointly is small (e.g. <10), this is not necessarily the case when analyzing a very large number of phenotypes. When considering a more diverse range of underlying models, involving multiple latent variables and various patterns of genetic effects, we found that increase in power can be achieved by applying a naive multistep approach, where signal on PCs are merged into sub-group based on their eigenvalues and the association signal across all groups tested jointly using the Fisher's method. This strategy and other approaches may be worth exploring in the future. While exploring these alternative models, we also noticed the consistent efficiency of MANOVA that achieved close to optimal power in many instances (**Figure S5**).

We simulated a wide range of trait correlation and pleiotropic patterns, nevertheless, no simulation study can be exhaustive. It is possible that analyses based on a specific mechanistic hypotheses may be more powerful when the mechanistic hypotheses hold. But such methods often lose power when the hypothesized mechanism does not hold. For example, the recently-proposed TATES statistic was shown to outperform MultiPhen in a range of scenarios when the genetic effect of a SNP was constant across multiple traits (e.g. $v_1=v_2$ in equations (A) and (B))[14]. When we compared the performance of tests in situations where the genetic effects varied across traits, TATES had notably less power than other approaches, including the combined PCs strategy (**supplementary figure S4** and **S5**). Although there exist situations where some tests may have more power, we believe the combined PC approach is an attractive approach as it retains relatively good power across a wide range of alternatives. This makes the combined PC approach particularly appealing when the underlying mechanism is unknown.

Finally, as for standard linear regression of the raw phenotypes, linear regression of PCs is sensitive to outliers and can make the test invalid. However additional simulations we conducted, showed that an increase in type I error rate only appears in the presence of a large number of outliers with values that are an order of magnitude larger than expected. Moreover, deviation from the null hypothesis would affect mostly genetic variants with very low frequencies (MAF<1%) and only have a small impact on common genetic variant (data not showed). We did not address adjustment for covariates in our simulations, however the regression-based context allow to add any covariates of interest using either a two steps strategy, where effect of covariates are regressed out at step 1 and residuals are used as outcome at step two, or by simply adding the covariate in the univariate test of the PCs. Since PCs are linear combination of the original variables, the effects of the covariates are simply projected on the PCs according to loadings of the phenotypes. While previous works has suggested that adding covariates in the model might be more powerful [23], it is equivalent to the two-step approach when the tested SNPs are not correlated with the covariates or interacting with the covariates [24].

The genetic variants identified in the MARTHA study at genome-wide significance level are mostly known variants. The association between the *ABO* gene and FVIII and vWF has been known for decades [25,26]; the two variants associated to aPTT (rs710446 and rs1801020) have been reported previously [19], and the two loci associated with ACVn, *LRP4* and *F5*, have already been described by Oudot et al [16]. Two new variants were identified by the combined PCs approach at genome-wide or nearly genome-wide significance level, the SNP rs183013917 that is mainly associated with FVIII and the SNP rs76854392 near gene *NRP1*, which is associated to both FVIII and vWF. Although these are potential candidates, especially the *NRP1* gene because of previous studies showing association with angiogenesis (e.g. [27]), they deserved further replication before being confirmed.

A number of studies have used principal component analysis for the multivariate analysis of correlated phenotypes. Most of them followed the standard strategy which consists in focusing on the few top PCs explaining most of total variance and removing those explaining a low amount of total variance. In this work, we show that contrary to this widespread practice, testing the top PCs only can be dramatically underpowered since PCs explaining a low amount of the total phenotypic variance can harbor a substantial part of the total genetic association. We also demonstrate that PCA-based strategies only achieve a moderate gain in power in the presence of positive pleiotropy, but have great potential to detect negative pleiotropy or genetic variants that are associated with a single trait highly correlated to others.

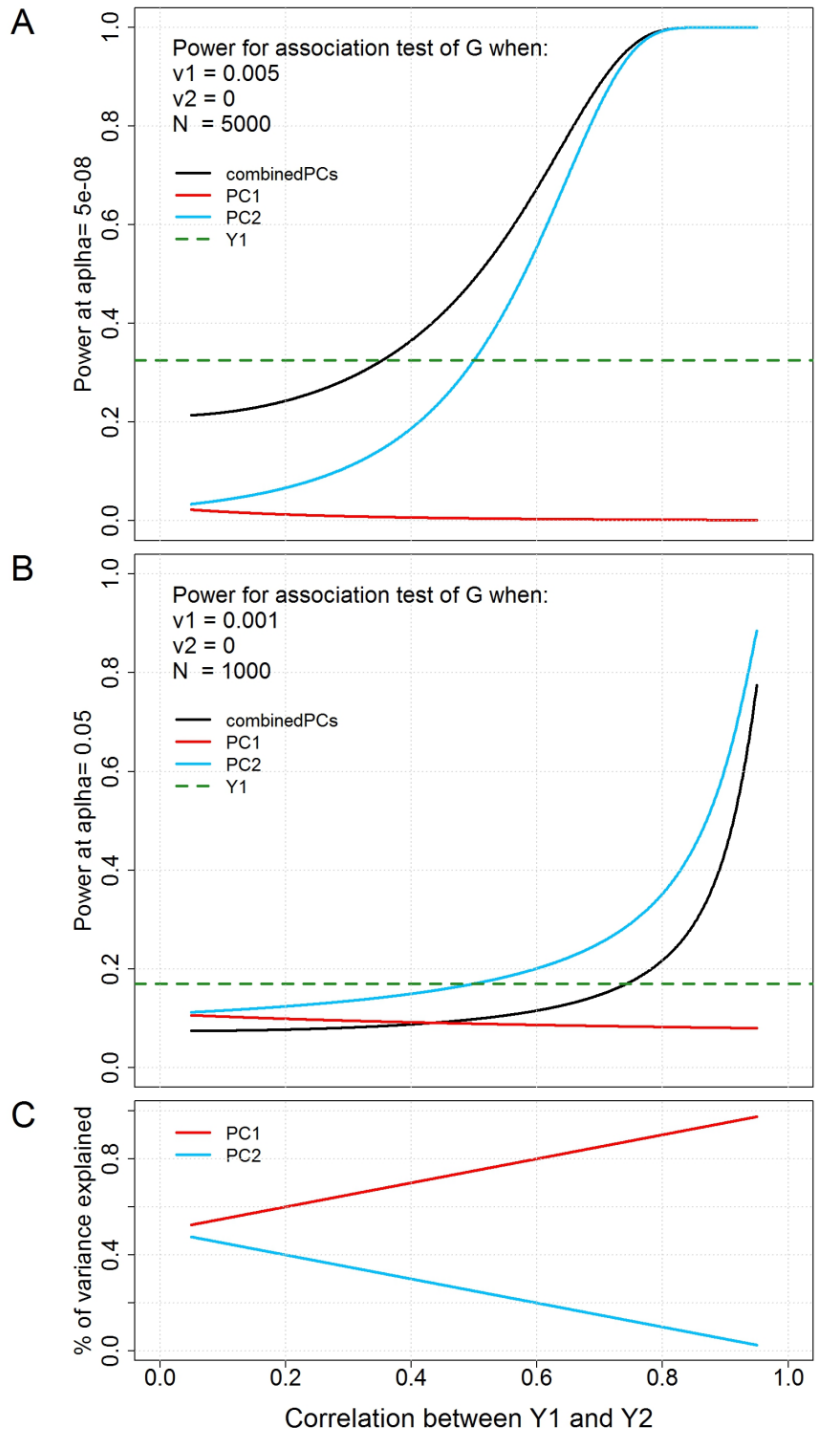

**Figure 1. Power to detect a SNP associated with a single trait in a bivariate analysis.** Power to detect the SNP associated with $Y_1$ based on the tests of $PC_1$, $PC_2$, the combined PCs, and $Y_1$ for different sample size and genetic effects (A and B), and proportion of variance explained by $PC_1$ and $PC_2$ (C). The power of each of the four tests is presented as a function of $c$ the correlation between $Y_1$ and $Y_2$, the sample size $N$, and $v_1$ and $v_2$, the proportion of the variance of $Y_1$ and $Y_2$ explained by the SNP respectively.

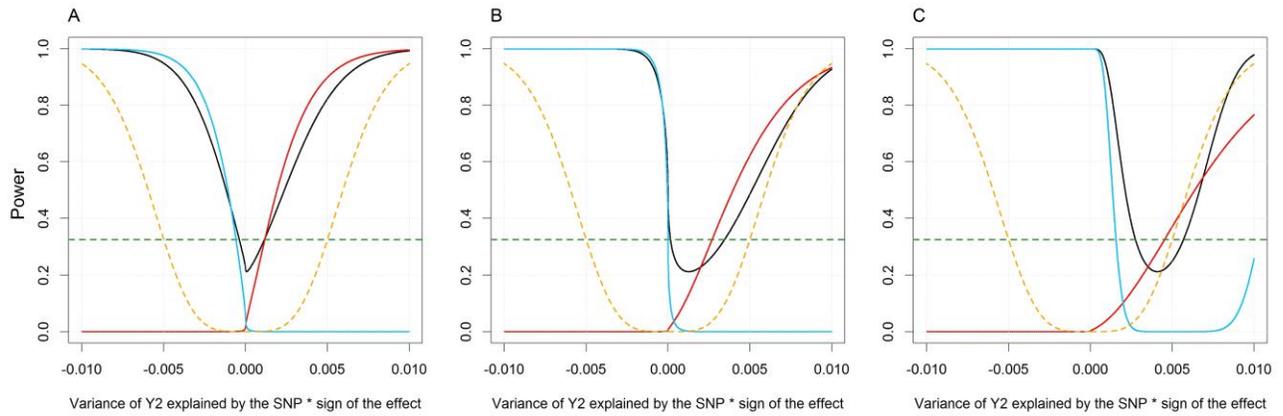

**Figure 2. Power to detect a SNP associated with two traits in a bivariate analysis.** Power at 5 x 10$^{-8}$ significance level to detect the SNP associated with the $Y$ using the independent tests of $Y_1$, $Y_2$, $PC_1$ and $PC_2$ and a combined PCs test when analyzing 5000 individuals. The genetic variant has a fixed effect on the trait $Y_1$. The power of each of the four tests is presented as a function of the effect on the second trait $Y_2$ for three level of correlation between $Y_1$ and $Y_2$, 0.1 (A), 0.5(B) and 0.9(C).

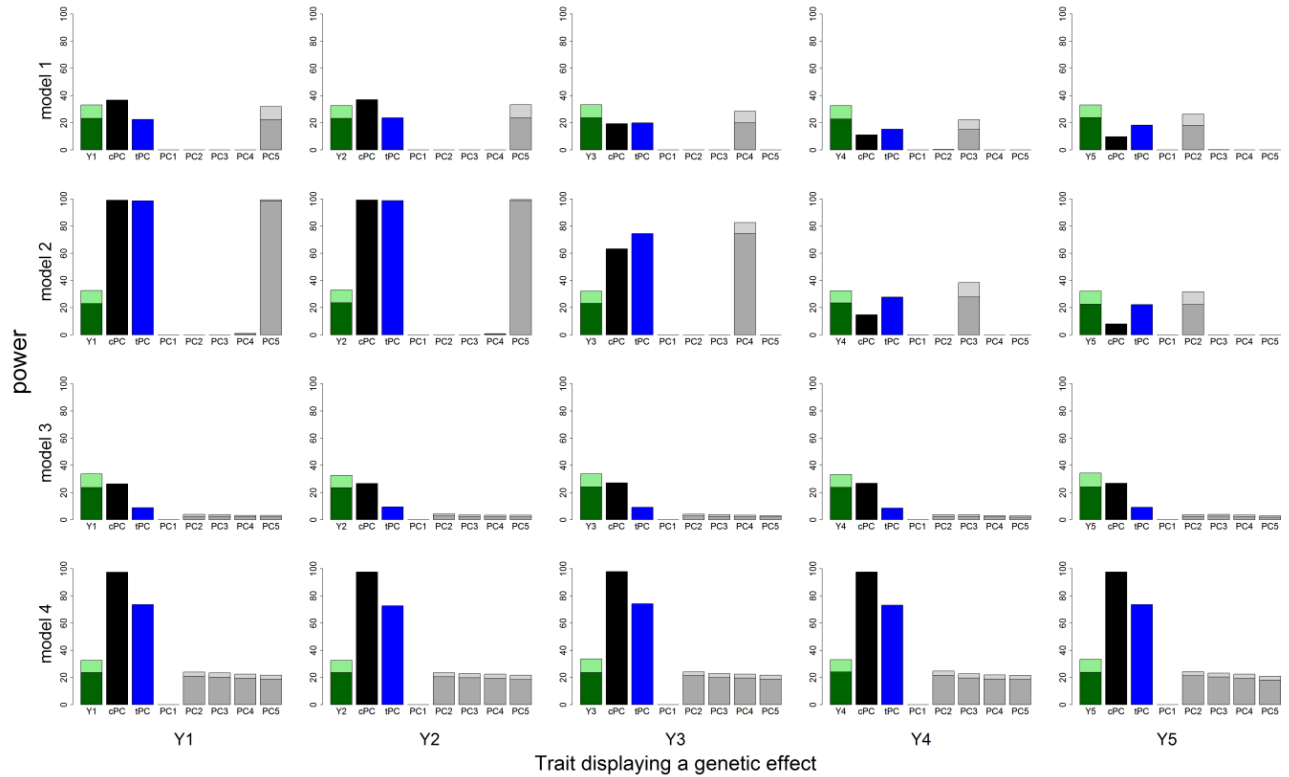

**Figure 3. Power comparison for the multivariate analysis of five traits in the absence of pleiotropic effect.**
Power at 5 x 10$^{-8}$ significance level for the detection of a genetic variant when analyzing five phenotypes. Only one phenotype is simulated as a function of the genetic variant, with proportion of variance explained equal to 0.5%. The bars represent the power of eight different tests. The univariate tests for each PC are shown in light grey and dark grey after correcting for the multiple testing, and the univariate test for the most significant PC (tPC) is in blue. The combined test of all five PCs is shown in black, and the univariate test for the associated Y in light green (dark green after correcting for the multiple testing). The power is shown for four different correlation models and 10,000 simulation replicates with 2000 individuals.

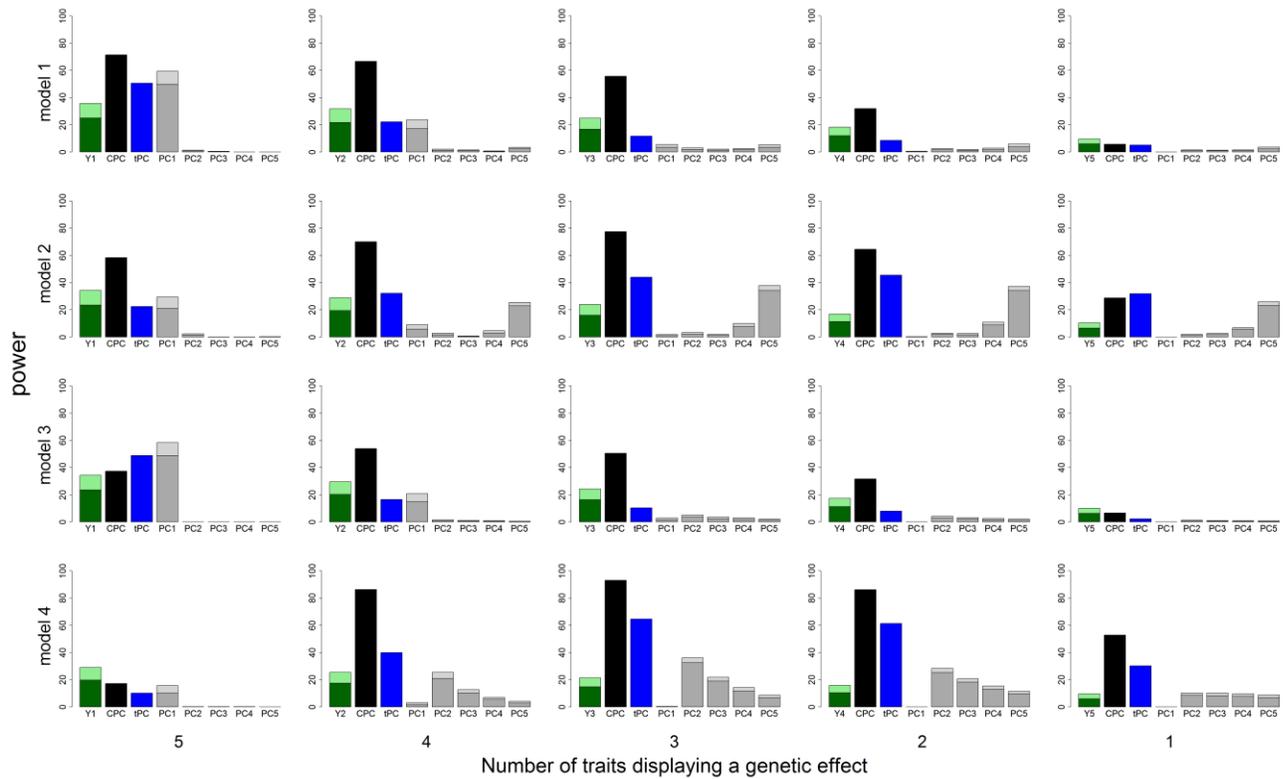

**Figure 4. Power comparison for the multivariate analysis of five traits in the presence of pleiotropic effect.** Power at 5 x $10^{-8}$ significance level for the detection of a genetic variant when analyzing five phenotypes. Between one and five phenotype are simulated as a function of the genetic variant, where its proportion of variance explained was randomly chosen between 0.1% and 0.5%. All genetic effects were positive and the associated phenotypes were randomly selected with equal probability. The bars represent the power of eight different tests. The univariate tests for each PC are shown in light grey and dark grey after correcting for the multiple testing, and the univariate test for the most significant PC (tPC) is in blue. The combined test of all five PCs is shown in black, and the univariate test for the corresponding Y in light green (dark green after correcting for the multiple testing). The power is shown for four different correlation models and 10,000 simulation replicates with 2000 individuals.

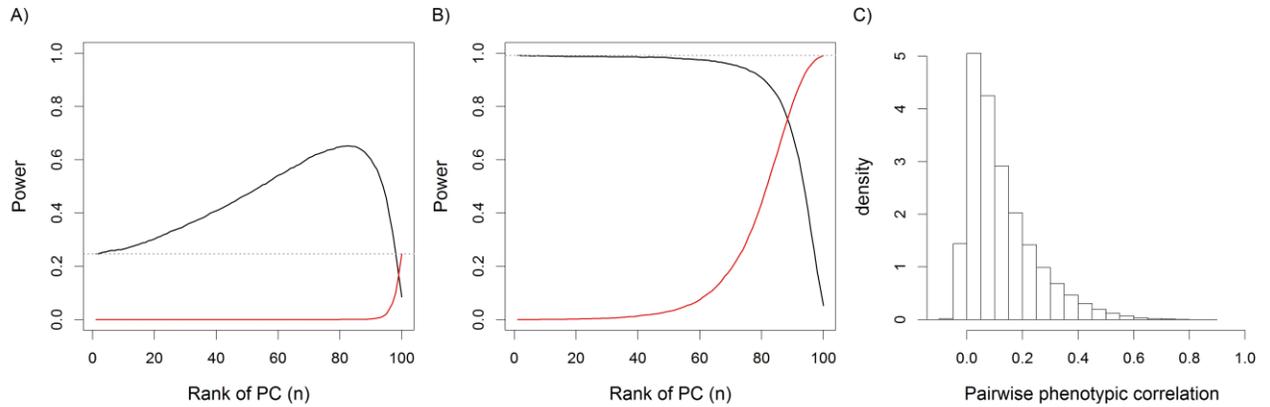

**Figure 5. Statistical power for multivariate analysis of 100 traits.** Power at 5 x 10$^{-8}$ significance level over 10,000 replicates for the detection of genetic variant when analyzing jointly 100 phenotypes in 2000 individuals. The correlations between the phenotypes were generated using 30 latent variables (see simulation scheme SC1 in supplement text S2 and figure S3), so that the pairwise phenotypic correlations follow a gradient from 0 to 0.9 (C). The red curve corresponds to the test combining the association signal from the first *n* PCs (largest eigenvalues), and the black curve corresponds to the test combining the association signal from the last 101-*n* PCs (smallest eigenvalues), with *n* varying from 1 to 100 and being equal to the rank of the PCs based on eigenvalues. The dashed grey line corresponds to the test that combines all PCs. Power is plotted when the genetic variant under study affect 5 phenotypes (A), 20 phenotypes (B). Phenotypes associated with the SNP are drawn with probability proportional to their correlation with other traits.

**Table 1. Rational for testing genetic association with PCs in a bivariate analysis**

| Test | PC1 | PC2 | Combined PCs |
|---|---|---|---|
| Genetic Model | | | |
| $\beta_{Y1} \neq 0$ and $\beta_{Y2} = 0$ | Almost no power, converging to 0 with increase correlation | Most powerful for small sample size and low $\beta_{Y1}$. Power increases with correlation | Most powerful for large sample size and large $\beta_{Y1}$. Power increases with correlation |
| $\beta_{Y1}$ in the same direction as $\beta_{Y2}$ | Most powerful when correlation and $\beta_Y$ are moderate. | Very low power. Power increase slightly with correlation | Most powerful when correlation and $\beta_Y$ are high. |
| $\beta_{Y2}$ opposite to $\beta_{Y1}$ | Almost no power. Minor variation with increase correlation | Very powerful. Power increase with correlation | Very powerful. Power increase with correlation |

*The two traits are denoted Y1 and Y2 and, and genetic effect of G on Y1 ($\beta_{Y1}$) and Y2 ($\beta_{Y2}$)*

Table 2. Genome-wide significant SNPs from the three approaches: univariate phenotypes, univariate PCs and combined PCs approach

| SNP id | Chr | Gene* | MAF | Rsq | P-value | | | | | | | | | | |
|---|---|---|---|---|---|---|---|---|---|---|---|---|---|---|---|
| | | | | | ACVn | FIB | FVIII | aPTT | vWF | PC1 | PC2 | PC3 | PC4 | PC5 | combPC |
| rs6025 | 1 | F5 | 0.90 | 0.93 | **5.3 x 10$^{-23}$** | - | - | - | - | 0.21 | 0.0091 | 1.1 x 10$^{-17}$ | - | - | 1.8 x 10$^{-18}$ |
| rs710446 | 3 | KNG1 | 0.54 | 0.99 | - | - | - | **5.2x 10$^{-11}$** | - | 0.064 | 0.0083 | - | 8.2 x 10$^{-8}$ | - | 3.3 x 10$^{-10}$ |
| rs1801020 | 5 | F12 | 0.78 | 0.86 | 0.29 | 0.19 | - | 2.2 x 10$^{-12}$ | - | - | 1.0 x 10$^{-5}$ | 0.0057 | 5.8 x 10$^{-9}$ | - | **7.6x 10$^{-15}$** |
| rs576123 | 9 | ABO | 0.55 | 0.86 | - | - | 2.7 x 10$^{-9}$ | 9.7 x 10$^{-5}$ | 1.1 x 10$^{-13}$ | 3.6 x 10$^{-10}$ | 0.00011 | - | 0.0062 | - | **1.2x 10$^{-14}$** |
| rs76854392 | 10 | NRP1 | 0.90 | 0.74 | - | - | 6.6 x 10$^{-7}$ | - | 0.0073 | 0.00075 | - | 0.029 | 0.28 | 0.00034 | **5.8x 10$^{-8}$** |
| rs191945075 | 11 | LRP4 | 0.96 | 0.60 | **2.8x 10$^{-25}$** | 0.095 | - | - | - | 0.017 | 0.00072 | 1.5 x 10$^{-17}$ | - | - | 1.3 x 10$^{-19}$ |
| rs183013917 | 18 | DCC | 0.99 | 0.82 | - | - | 1.7 x 10$^{-7}$ | - | 0.035 | 0.0017 | - | - | 0.12 | 7.0 x 10$^{-6}$ | **1.8 x 10$^{-8}$** |

*Abbreviation: Chr, chromosome; MAF, minor allele frequency; Rsq, quality control imputation criterion ; FIB, Fibrinogen ; FVIII, factor VIII ; vWF, von Willebrand factor; aPTT, the activated partial thromboplastin time ;  and ACVn, the standardized Anticoagulant response to Agkistrodon contortrix venom ;  combPC, combined PCs approach*

*physically closest gene.

The p-values for univariate raw phenotype analysis and univariate PC analysis were adjusted for multiple tests using the Bonferroni correction. P-values greater than 0.3 were replaced by the sign "-". The most significant test is indicated in bold.

# Supplementary material

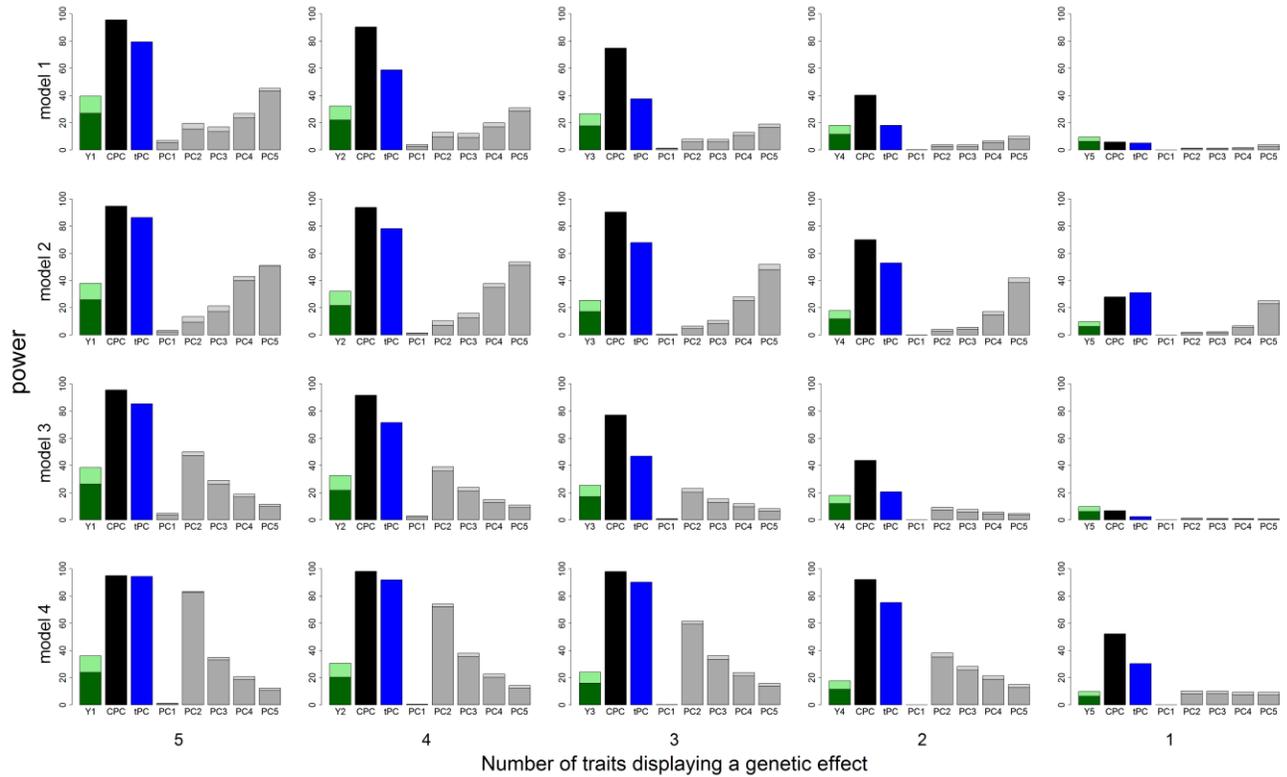

**Figure S1. Power comparison for the multivariate analysis of five traits in the presence of negative pleiotropy.** Power at 5 x $10^{-8}$ significance level for the detection of a genetic variant when analyzing five phenotypes. Between one and five phenotype are simulated as a function of the genetic variant, where its proportion of variance explained was randomly chosen between 0.1% and 0.5%. The direction of the genetic effects was randomly chosen for each simulation and the associated phenotypes were randomly selected with equal probability. The bars represent the power of eight different tests. The univariate tests for each PC are shown in light grey and dark grey after correcting for the multiple testing, and the univariate test for the most significant PC (tPC) is in blue. The combined test of all five PCs is shown in black, and the univariate test for the corresponding Y in light green (dark green after correcting for the multiple testing). The power is shown for four different correlation models and 10,000 simulation replicates with 2000 individuals.

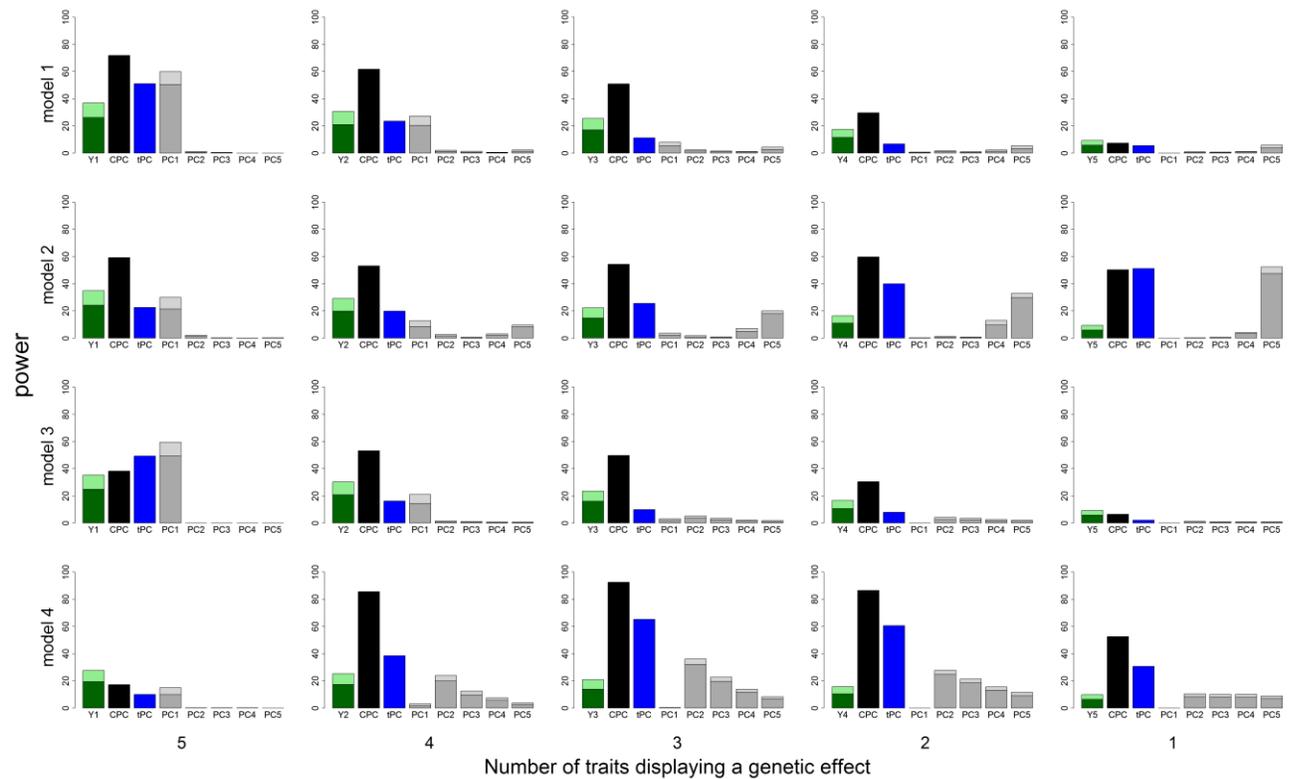

**Figure S2. Power comparison for the multivariate analysis of five traits when pleiotropic effects depend on correlation.** Power at $5 \times 10^{-8}$ significance level for the detection of a genetic variant when analyzing five phenotypes. Between one and five phenotype are simulated as a function of the genetic variant, where its proportion of variance explained was randomly chosen between 0.1% and 0.5%. All genetic effects were positive and associated phenotypes were selected randomly with probability proportional to their level of correlation with other phenotype. The bars represent the power of eight different tests. The univariate tests for each PC are shown in light grey and dark grey after correcting for the multiple testing, and the univariate test for the most significant PC (tPC) is in blue. The combined test of all five PCs is shown in black, and the univariate test for the corresponding Y in light green (dark green after correcting for the multiple testing). The power is shown for four different correlation models and 10,000 simulation replicates with 2000 individuals.

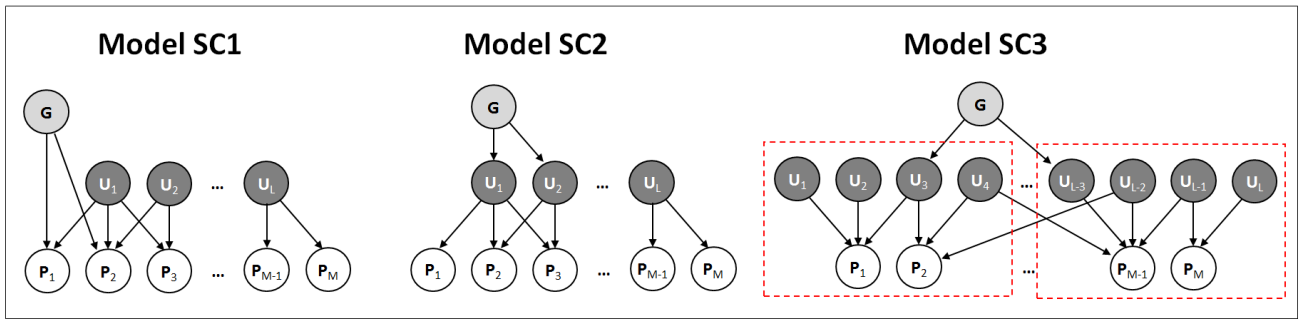

**Figure S3. Examples of simulation schemes.** An illustrative example of the three hypothetical simulation schemes considered when generating a large number of correlated phenotypes. In model SC1 and SC2, the number of latent variable *L* is much smaller than the number of phenotypes M, while in model SC3, *L* is much larger than *M*. In model SC3, the phenotypes are grouped into clustered defined as function of sets of latent variables (red frame), although some latent variable can be associated with phenotype in different clusters.

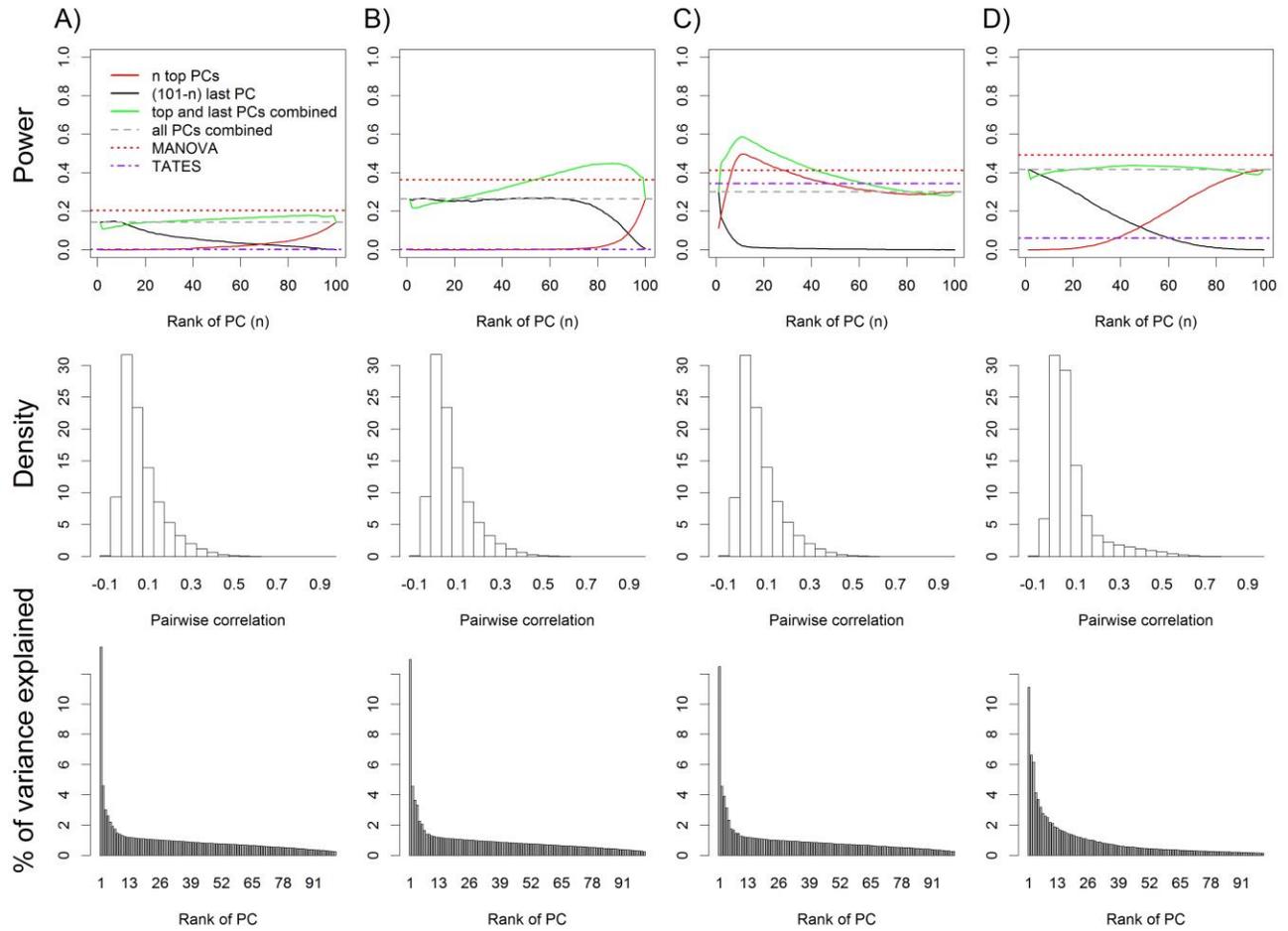

**Figure S4. Power comparison for the multivariate analysis of 100 traits**: Plots from column A), B),C), and D) were simulated under simulation schemes SC1, SC1, SC2 and SC3 respectively (the schemes are describe in text S2 and figure S3). The top panel shows the power at 5 x 10$^{-8}$ significance level for six different tests. The plain red curve corresponds to the test combining signal from the *n* PCs associated with the largest eigenvalues, the plain black curve corresponds to the test combining information on the 101-*n* PCs associated with the smallest eigenvalues, and the green curve correspond to the combined test of two previous tests using the Fisher's method, with *n* varying from 1 to 100. The dashed grey line corresponds to the test of all PCs combined, the dashed orange line to MANOVA and the dashed purple line to the TATES method. The middle panel displays the distribution of pairwise phenotypic correlation and the bottom panel distribution of the proportion of variance explained by each PC ordered by eigenvalues when analyzing 100 phenotypes.

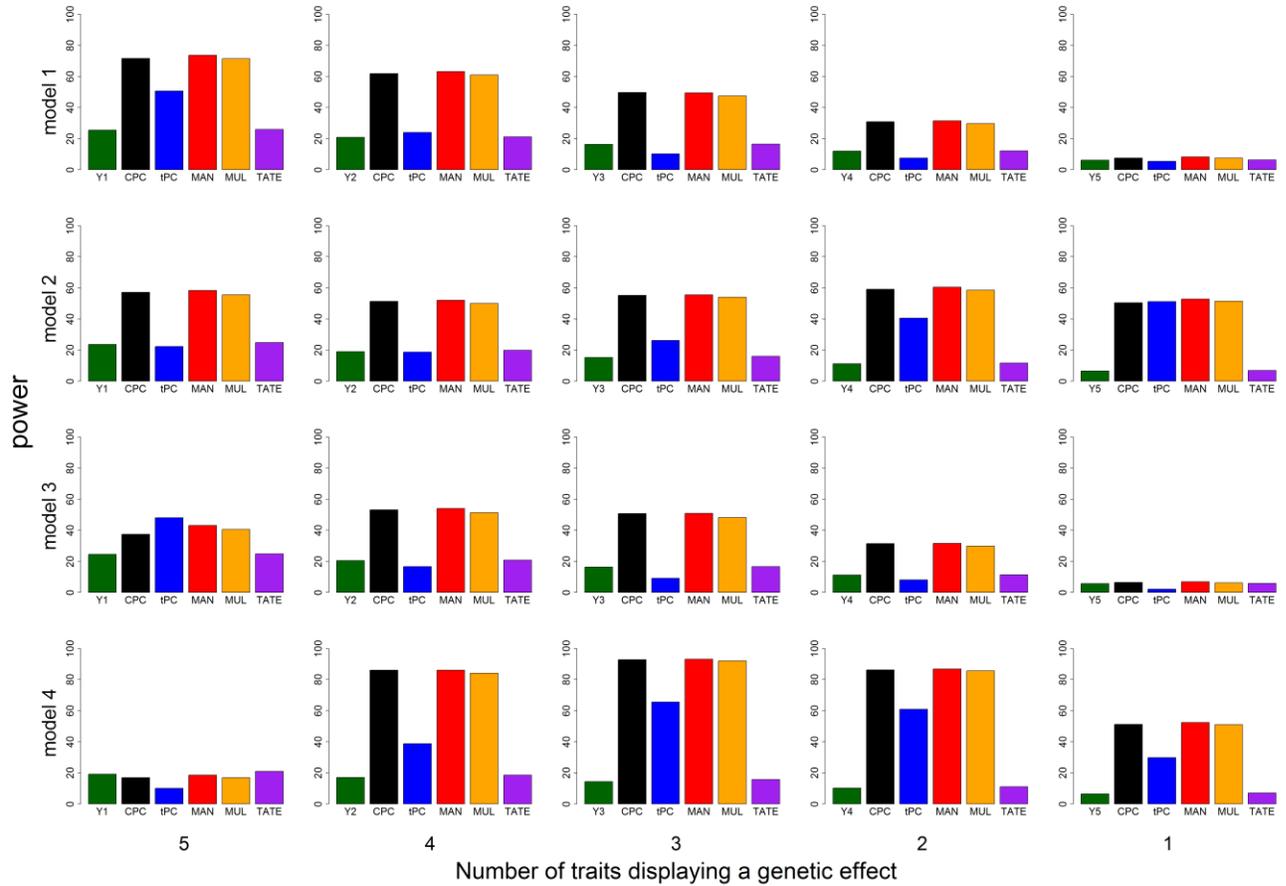

**Figure S5. Power of alternative methods for the multivariate analysis of five traits.** A comparison of the power (at 5 x 10$^{-8}$ significance level) for detecting a genetic variant using a standard univariate test (green) and five different multiple trait analysis: most significant PC (tPC, in blue); the combined PCs (in black); Manova (MAN, in red); Multiphen (MUL, in orange); TATES (TATE, in purple). These test were applied to five traits where the number of traits with causal genetic effect was varied between one and five. The five traits were simulated under simulated under four different correlation models, each with 10,000 replicates and 2000 individuals. The proportion of variance explained by the genetic variant explained was randomly chosen between 0.1% and 0.5%. All genetic effects were positive and associated phenotypes were selected randomly with probability proportional to their level of correlation with other phenotype.

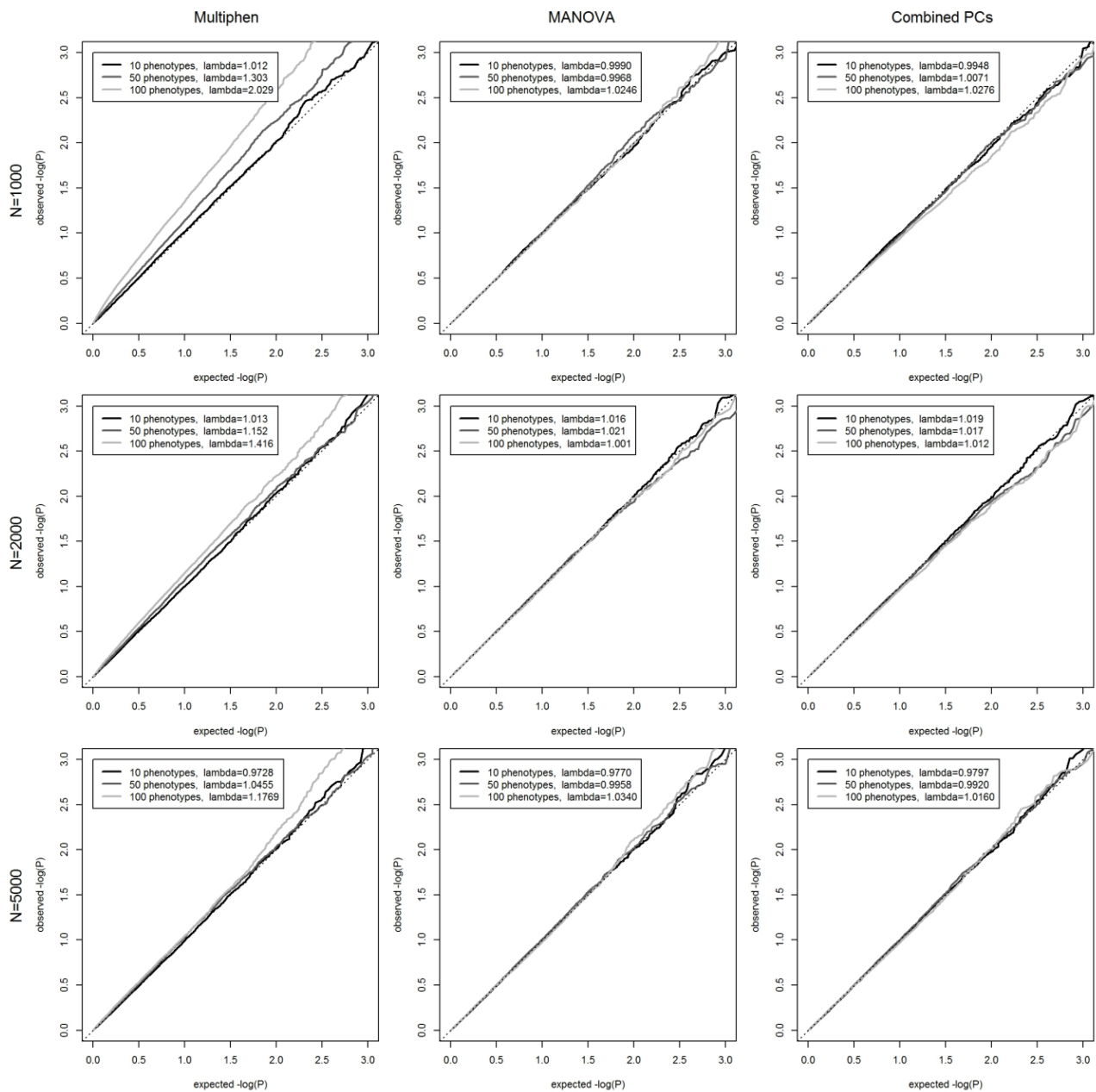

**Figure S6. QQplot of Multiphen, MANOVA and the combined PC approach under the null.** : QQ-plots and the inflation factor (lambda) under the null hypothesis of no association between the tested genetic variant and any of the phenotype under study, when analyzing 10, 50 and 100 phenotypes in 1000, 2000 and 5000 individuals among 10,000 replicates for Multiphen (left column), MANOVA (central column) and, the combined PCs approach (right column), when the correlations between phenotypes is following a gradient from 0 to 0.8.

**Table S1.** Correlation matrix between the five coagulation factors related phenotypes

|       | FVIII  | vWF    | aPTT   | ACVn   | FIB    |
|-------|--------|--------|--------|--------|--------|
| FVIII | 1      | 0.758  | -0.452 | -0.333 | 0.243  |
| vWF   | 0.758  | 1      | -0.290 | -0.227 | 0.232  |
| aPTT  | -0.452 | -0.290 | 1      | 0.245  | -0.013 |
| ACVn  | -0.333 | -0.227 | 0.245  | 1      | -0.055 |
| FIB   | 0.243  | 0.232  | -0.013 | -0.055 | 1      |

*Abbreviations: FIB, Fibrinogen ; FVIII, factor VIII ; vWF, von Willebrand factor; aPTT, the activated partial thromboplastin time ; and ACVn, the standardized Anticoagulant response to Agkistrodon contortrix venom.*

**Table S2.** Loadings and proportion of variance explained for each principal component

| Original phenotypes | Principal components | | | | |
|---|---|---|---|---|---|
|                | *PC1*  | *PC2*  | *PC3* | *PC4*  | *PC5*  |
| ACVn           | 0.248  | -0.253 | 0.931 | 0.052  | 0.064  |
| FVIII          | -0.599 | -0.107 | 0.092 | -0.238 | 0.751  |
| vWF            | -0.561 | -0.045 | 0.208 | -0.488 | -0.634 |
| FIB            | -0.279 | 0.834  | 0.281 | 0.384  | -0.016 |
| aPTT           | 0.432  | 0.476  | 0.044 | -0.745 | 0.172  |
| %var explained | 0.462  | 0.189  | 0.183 | 0.125  | 0.041  |

*Abbreviations: FIB, Fibrinogen ; FVIII, factor VIII ; vWF, von Willebrand factor; aPTT, the activated partial thromboplastin time ; ACVn, the standardized Anticoagulant response to Agkistrodon contortrix venom ; %var explained, proportion of total phenotypic variance explained.*

**Table S3.** Genomic inflation factor ($\lambda$) for the univariate and multivariate analysis

| Phenotype    | $\lambda$ |
|--------------|--------|
| ACVn         | 1.0117 |
| FIB          | 0.9991 |
| FVII         | 0.9865 |
| aPTT         | 1.0644 |
| vWF          | 0.9865 |
| PC1          | 0.9907 |
| PC2          | 0.9958 |
| PC3          | 1.0009 |
| PC4          | 1.0065 |
| PC5          | 1.0056 |
| Combined PCs | 0.9939 |

*Abbreviations: FIB, Fibrinogen ; FVIII, factor VIII ; vWF, von Willebrand factor; aPTT, the activated partial thromboplastin time ; ACVn, the standardized Anticoagulant response to Agkistrodon contortrix venom ; %var explained, proportion of total phenotypic variance explained.*

**Text S1**

The proportion of the total variance explained by $PC_1$ and $PC_2$, respectively $s_1$ and $s_2$, is defined as:

$$s_1 = \frac{var(PC_1)}{var(Y_1) + var(Y_2)}$$

$$= \frac{var(PC_1)}{2}$$

$$= \left(\frac{4*c}{2} + \frac{(\sqrt{v_1} + \sqrt{v_2})^2}{2} + \frac{(1-c-v_1)}{2} + \frac{(1-c-v_2)}{2}\right)/2$$

$$= \left(\frac{4*c + v_1 + v_2 + 2*\sqrt{v_1 v_2} + 1 - c - v_1 + 1 - c - v_2}{2}\right)/2$$

$$= \left(\frac{2 + 2*c + 2*\sqrt{v_1 v_2}}{2}\right)/2$$

$$= (1 + c + \sqrt{v_1 v_2})/2$$

$$s_2 = \frac{var(PC_2)}{var(Y_1) + var(Y_2)}$$

$$= \frac{var(PC_2)}{2}$$

$$= \left(\frac{(\sqrt{v_1} - \sqrt{v_2})^2}{2} + \frac{(1-c-v_1)}{2} + \frac{(1-c-v_2)}{2}\right)/2$$

$$= \left(\frac{v_1 + v_2 - 2*\sqrt{v_1 v_2} + 1 - c - v_1 + 1 - c - v_2}{2}\right)/2$$

$$= \left(\frac{2 - 2*c - 2*\sqrt{v_1 v_2}}{2}\right)/2$$

$$= (1 - c - \sqrt{v_1 v_2})/2$$

The proportion of variance of $PC_1$ and $PC_2$ explain by $G$, respectively $v_{pc1}$ and $v_{pc2}$ can be expressed as a the ratio of the genetic effect of $G$ on the variance of $PC_1$ and $PC_2$, that we denoted $\tau_1$ and $\tau_2$, respectively, divided by the total variance of $PC_1$ and $PC_2$.

$$v_{pc1} = \frac{\tau_1}{var(PC_1)}$$

$$= \frac{\left(\frac{\sqrt{v_1} + \sqrt{v_2}}{\sqrt{2}}\right)^2}{(1 + c + \sqrt{v_1 v_2})}$$

$$= \frac{v_1 + v_2 + 2*\sqrt{v_1 v_2}}{2*(1 + c + \sqrt{v_1 v_2})}$$

$$v_{pc2} = \frac{\tau_2}{var(PC_2)}$$
$$= \frac{\left(\frac{\sqrt{v_1}-\sqrt{v_2}}{\sqrt{2}}\right)^2}{(1-c-\sqrt{v_1 v_2})}$$
$$= \frac{v_1 + v_2 - 2*\sqrt{v_1 v_2}}{2(1-c-\sqrt{v_1 v_2})}$$

**Text S2**

When simulating 100 phenotypes, we considered three different simulation schemes, SC1, SC2 and SC3, which are illustrated in **figure S3**. In model SC1 the correlation between phenotypes is a result of a limited number (30) of independent latent variables, each affecting 40 phenotypes on average and explaining altogether 30% of the total phenotypic variance. The genetic effect on the phenotypes was generated independently from the latent variables, accounting or not for the pairwise phenotypic correlation. In model SC2, the phenotypic correlation was generated in similar fashion as in model SC1, i.e. using the same latent variables; however the genetic variant was associated with some of these latent variables but not directly to the phenotypes. In model SC3, we considered a more complex model involving thousands of latent variables together explaining 90% of the total phenotypic variance, sub-groups of these latent variables affecting cluster of phenotypes. As for model SC2, the genetic variant was associated with some of the latent variables.

More specifically, in model SC1, the $n_{phe}$ phenotypes were generated as follows: $\mathbf{y} = \sqrt{\mathbf{c}} \times (\boldsymbol{\beta}^t \times \mathbf{u}) + \sqrt{\boldsymbol{\gamma}} \times g + \sqrt{(1 - \boldsymbol{\gamma} - \mathbf{c})} \times \boldsymbol{\varepsilon}$, where **y** is a vector of phenotypic values for a given subject, g is a SNP and $\boldsymbol{\gamma}$ the vector of proportion of variance explained by that SNP on the $n_{phe}$ phenotypes; **u** is a vector of realization of $n_u$ independent latent variables $U = (U_1, \dots U_{n_u})$ normally distributed with mean 0 and variance 1, $\boldsymbol{\beta}$ is a matrix of (positive) weights with $n_u$ rows and $n_{phe}$ columns that defines the contribution of each variable $U_i$ to the phenotypes, these weights are defined so that for each phenotype $j$, $\sum_{i=1}^{n_u} \beta_{ij}^2 = 1$; **c** is a vector of $n_{phe}$ weights that defined the proportion of variance explained by each linear combination $\boldsymbol{\beta}_i^t \times U_i$ ; and $\boldsymbol{\varepsilon}$, the residual variance is a vector of $n_{phe}$ independent variables, normally distributed with mean 0 and variance 1. In model SC2 and SC3 the $n_{phe}$ phenotypes were generated as follows: $= \sqrt{\mathbf{c}} \times (\boldsymbol{\beta}^t \times \mathbf{u}) + \sqrt{(1 - \mathbf{c})} \times \boldsymbol{\varepsilon}$. A sub-sample of the $U_i$ were generated as a function of a SNP G such that $U_i = \sqrt{\delta_i} \times g + \sqrt{(1 - \delta_i)} \times e_i$, where $\delta_i$ is the effect of SNP on the latent variable $U_i$, and $e_i$, the residual, is normally distributed with mean 0 and variance 1.

The parameters **c** and **β** and **γ** of the three models were defined empirically to obtain a similar distribution of pairwise phenotypic correlation and a similar distribution of proportion of variance explained by the PCs (**figure S4**). Since the three models were very different, it was difficult to generate a similar range of genetic effects on the phenotypes. However, this was of secondary importance as the aim of this experiment was not to compare the models but to compare the power of different methods for analyzing correlated traits. Regardless, with these constraints, the genetic effects simulated (either directly on the phenotypes or on the latent variables) were explaining a very small proportion of the total phenotypic variance (<0.05%). For example, in model SC1, the SNP was associated with 10 traits and the variance that it explained for each phenotype was drawn from a uniform distribution with minimum 0.1% and maximum 0.5%, so that the total phenotypic variance explain was on average 0.025%. In model SC2, the SNP had a similar effect size on 3 latent variables, so that the average contribution of the SNP was 0.007%. These parameter choices resulted in similar distributions of pairwise phenotypic correlations (see figure S4).